# RAGLog: Log Anomaly Detection using Retrieval Augmented Generation

Jonathan Pan, Swee Liang Wong, Yidi Yuan
Home Team Science and Technology Agency, Singapore
Jonathan_Pan@htx.gov.sg, Wong_Swee_Liang@htx.gov.sg, Yuan_Yidi@htx.gov.sg

*Abstract*— The ability to detect log anomalies from system logs is a vital activity needed to ensure cyber resiliency of systems. It is applied for fault identification or facilitate cyber investigation and digital forensics. However, as logs belonging to different systems and components differ significantly, the challenge to perform such analysis is humanly challenging from the volume, variety and velocity of logs. This is further complicated by the lack or unavailability of anomalous log entries to develop trained machine learning or artificial intelligence models for such purposes. In this research work, we explore the use of a Retrieval Augmented Large Language Model that leverages a vector database to detect anomalies from logs. We used a Question and Answer configuration pipeline. To the best of our knowledge, our experiment which we called RAGLog is a novel one and the experimental results show much promise.

## I. Introduction

The analysis of logs to detect anomalies is an important research topic with practical importance in the field of failure identification [1], [2] and security threat detection [3], [4]. Logs are generated by systems or applications that are codified and configured to report relevant information about the state of the applications or software while running. Here, the application may refer to any software running to perform specific task or tasks. It could be a mobile application, operating system running inside an Internet of Things (IoTs) device or a cloud compute node performing computational tasks. It could also be an environment of compute nodes working collectively on multiple tasks. Such logs and their log entries are generated based on its current configuration at the time of the log generation. These entries are also affected by the state of the application during its execution and its dependent factors that may originate from within the operating environment and executing platform of the application. It would be affected by external factors like users or external systems interacting with the application.

These internal and external factors affecting the log generation may change abruptly and progressively over time resulting in corresponding log entries being included into the log generation process. These factors may originate from planned changes like planned maintenance tasks. They may also originate from unplanned activities. Additionally, these changes may be induced by benign or malicious intent. For the latter, with the intent to evade detection, even if the logs are not tampered, its entries will be elusive to classical detection techniques. These further complicates the composition of logs to be analyzed.

The objective of performing analysis on logs is done to facilitate the detection of anomalous activities so that immediate or corresponding remediation may be done to contain or remediate the issue recorded in the logs. This is part of the attempt to enhance system resiliency against system faults, degradation and intentionally induced cyber physical attacks. It is also used to facilitate the investigation or analysis of what may have induced the occurrence of such anomalous activities. The scope of this research work is on the detection of such anomalous activities from the logs. However, due to the characteristics of logs, namely being voluminous, varied, and contextual, regular log analysis is difficult, warranting the need for automation. While rule or signature-based automation solution helps, the contextual or semantic complexity of logs limits its efficacy [16].

There is many research work done to develop AI algorithms to detect anomalies from logs. However, log analysis using AI algorithms require extensive preparation and data requirements to train the models [6][7]. The capability of log analysis using AI algorithms to detect anomalies has several challenges and constraints to deal with before it contributes significantly to its intended objectives of keeping system resilient. For supervised models, there is the challenge of acquiring sufficient anomalous data points to train such models. For unsupervised models, it will be the ability to detect the variety and variations of anomalies in logs. Recent research work to apply Large Language Models (LLMs) to log analysis processes have shown promising results but are constrained by model limits such as token capacity, ability to remember and hallucinations. Also, there is limited evaluation on its efficacy.

In our research work, we seek to address these constraints using a Retrieval Augmented Generation approach with a vector database and evaluate its performance in detecting log anomalies. Our novel log anomaly detection solution termed RAGLog uses a Retrieval Augmented Generation construct with a vector database to store subset of normal log entries and a Large Language Model to perform zero shot semantic analysis of the queried log entry. Our pipeline process requires minimal data preprocessing and does not require log parsing. It uses unsupervised clustering to enhance log anomaly detection.

In the next section, we will cover the challenges and complexity of performing log analysis. This is followed by a review of current log analysis algorithms including Large Language Models. A coverage of our RAG construct is described in the section that follows with the details of the experimental setup and its evaluation. This paper concludes with a summary of this work and potential future research direction.

## II. BACKGROUND INFORMATION

In this section, we articulate the background information related to the need for the analysis of logs, its complexity, and challenges with current log analysis algorithms.

### A. Need for Analysis of Logs

Logs are generated by software driven applications running on systems or devices to provide information to aid developers and system engineers with their analysis of system's state and condition. It is also used as a form of audit trail to log the occurrences of events in chronological manner. The analysis of logs is also done to facilitate investigation after the occurrence of an incident related to the system that generates the logs. This incident could be in the form of system defect and a malicious or unintended breach of the system. With investigation, the log could provide the means to reconstruct the occurrence of the incident. With such forms of analysis, an investigator or system engineer would attempt to identify the occurrence of anomalous events through the logs. However, to identify such anomalies, one would need to know how to spot such anomalies from voluminous entries posted into the log files.

### B. Challenges with Log Analysis

The form for logs is typically unique to how the software has been developed or configured to post entries into these textual files. Also, each system or software component may adopt its own logging format and information lexicon representation that details the state of the run-time system when logs are posted. Such information within the logs is highly context specific to the environment which the system resides in [16]. For example, information like the IP addresses or hostnames or resource identities. Entries in the logs are dependent also on the configuration surrounding the involved system and their own respective environmental conditions. In addition to the contextual settings, the log entries are sequenced by its chronological occurrence of events or state. Hence such log entries have a time dimension.

Hence, the analysis of such log datasets requires contextual understanding of the system or component that generates such logs. Also, the analysis requires the means to classify or distinguish what is a normal log entry and what is not a normal log entry. For the latter, such information of recognizing an abnormal entry would be constrained to what may be conceivable based on the engineering design of the system involved or known instances of events that could cause an anomalous event like a cyber security breach attempt. However, there will be instances where such information or knowledge is only acquired through the occurrence of the event that in turn induces the anomalous log entries. Hence the challenges with log analysis are the need for semantic comprehension [15][16] to perform good log analysis and the challenge of having limitedly available information about the form of anomalies that could occur.

### C. Challenges with Large Language Models

Current Large Language Models (or Generative Artificial Intelligence) have inherent limitations that includes limits to the size of the tokens that they can handle which in turns limits how much contextual information LLMs can take in or recall as well as potential for hallucinations [17]. Solutions are being researched upon to address such limitations with information retrieval techniques that will be described in subsequent sections.

## III. RELATED WORK

In this section, we review the current log analysis algorithmic development and their strengths and limitations.

### A. Multi-staged Log Processing

The current log analysis algorithmic designs typically involve multiple stages of log processing before analysis is applied. It typically starts with log parsing that converts raw logs into structured data features. These extracted features would undergo further transformation as they are typically represented as textual features and would be converted to numerical forms. Log partitioning typically follows that involves converting the contiguous log into associative partitions to improve anomaly classification. This may involve the use of time-based partitions, partitions organized by windows of similar or compatible operations or identifier-based divisions of log entries. Finally, the anomaly detection algorithm would then be applied after these pre-processing.

Thus far, there are very few developmental attempts to develop an integrated model that could ingest raw log data for immediate model training and inference. Based on our survey, one by Hashemi and Mäntylä [5] and Le and Zhang [15] ingress logs without log parsers. Le and Zhang observed that log parsers could cause inaccurate log parsing due to misinterpretation of the semantic meaning of the log analysis and not handle Out-of-vocabulary (OOV) words well. Our approach removes the need for log parsing, allowing inferences on raw log data inputs.

### B. Algorithms to detect Anomalous Events from Logs

Many of the log analysis algorithms focused on the key area of detecting anomalous events from logs. From the survey work done by He et al. [6] and Chen et al. [7], the algorithms are either supervised or unsupervised machine learning algorithms. These algorithms may be based on classical machine learning algorithms or deep learning algorithms. Supervised learning algorithms are constrained by the availability of anomalous data with labels within the training datasets. Additionally, even with the availability of anomalous data within the log datasets, the class imbalance could pose a significant challenge to the training of the model. Also, these models may need to undergo retraining or be reconstructed to internalize the new knowledge. With unsupervised learning algorithms [12][13] for log analysis, their challenge is the efficiency of the algorithms to detect the variety and variations of anomalies captured in log entries as such anomalies may occur and vary significantly over a prolonged period of time. When new anomalies are discovered, these unsupervised models will require retraining. Our approach uses the vector database to store only small samples of normal log entries. The LLM will do anomaly detection without any samples of anomalous log entries. Hence it is a zero shot classifier.

*C. LLM for Log Analysis*

There were recent attempts to apply Large Language models to perform log analysis. Qi et al. [18] proposed a framework for log-based anomaly detection using ChatGPT using varied prompt constructs, window sizes and input sequences. Their work showed the non-triviality of an optimal prompt, window size limitations as well as high false positive rates. Mudgal et al. [19] designed specific prompts with ChatGPT for log parsing that had excellent performance. However, with other areas of log analysis like anomaly detection and log summarization, the LLM exhibited limitations that warrant further research. Liu et al. [20] tested their LogPrompt model in zero-shot scenarios with varying number of provided log samples and different prompt formats (self-prompt, CoT prompts and In-context Prompt). The zero-shot test results showed promise when compared with our log analysis algorithms and other Deep Learning architectures. However, it had very low precision scores which is not optimal if applied to log analysis for operations and maintenance activities to support resiliency.

These preliminary experimentations demonstrate the necessity for further research in applying LLM for log analysis, especially in detecting anomalies in logs, forming the basis of this research work.

## IV. MODEL

Our construct uses the Retrieval Augmented Generative (RAG) model [22] to analyze log entries by querying its store of samples of normal log entries. In our work, the store is a vector database. The Large Language Model would need to perform semantic analysis between the retrieved log samples from the database and the queried log entry.

This creates an end-to-end model construct that is simple to use and adept for any log source, unlike many other similarly purposed algorithms mentioned in our previous section for log analysis which require the use of multi-stage log processing pipeline.

*A. Formulation for RAG*

The RAG operates in two stages. The first is the retrieval of contextually relevant information. The second involves using the retrieved information to generate the corresponding response. This can be formulated with $x$ as the provided input, which is the queried log entry, $z$ as the set of relevant log entries from the vector database and $y$ as the generated output from the LLM $f$. This can be expressed in the form.

$$y = f(x, z) \qquad (1)$$

Our model construct uses dense-vector retrieval approach [23] that encodes the log entries into vector embedding representations using a pre-trained Embedding model from OpenAI. The retrieval score is computed through inner products between the queried vector from the provided log entry against vectors stored in the vector database that contains provided samples of normal log entries. When the retrieval score matches the criteria like highest similarity score or minimal threshold score, the vector database will return the resultant vectors. The retriever that we used is from LangChain [24].

*B. LLM for Semantic Analysis*

With the retrieved vectors, the vector embedding will be decoded using the corresponding decoder by Embedding model. This turns the vector back to the original log entry representation. We frame the log anomaly detection as a Question and Answer [23], using a Question and Answer prompt template to include the best matched retrieved normal log entries to analyze whether a queried log entry is normal or abnormal. The prompt is fed to the Language Large Model.

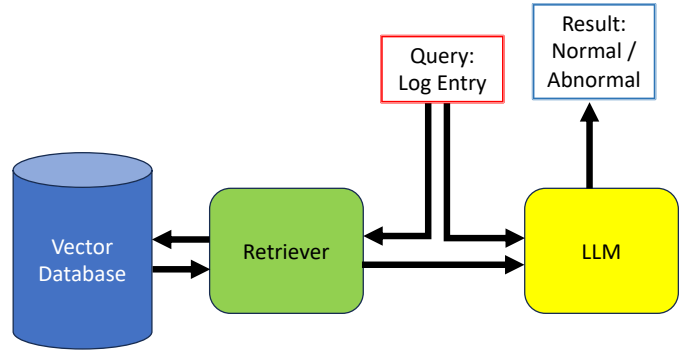

Fig. 1. RAGLog Architecture

## V. METHODOLOGY AND ANALYSIS

In our experiment setup, we designed our experiment to address our research question whether Retrieval Augmented Generation in LLM could perform log anomaly detection.

*A. Log Datasets*

For our log datasets, we used BGL [9] and Thunderbird [21]. These are two popular datasets typically used by researchers to evaluate their log models [7].

The BGL are open real-world datasets from HPC from a BlueGene/L supercomputer at Lawrence Livermore National Labs. This dataset has an important characteristic associated with their appearance of many new log messages in the timeline of the data, that is, the systems change over time. The Thunderbird open dataset of logs was collected by Sandia National Lab. It contains alert and non-alert messages. Both datasets are labelled with sizeable imbalance for the anomaly class.

*B. Evaluation Metrics*

As the dataset used had binary classification labels, we used *Precision* to measure the accuracy of the model against type I error (true positive) and *Recall* to measure the accuracy of the models against type II error (true negative). Finally, we used *F1 score* to measure the harmonic mean of *precision* and *recall*.

$$Precision = \frac{TP}{TP+FP} \quad (2)$$

$$Recall = \frac{TP}{TP+FN} \quad (3)$$

$$F1\ score = 2 \times \frac{Precision \times Recall}{Precision + Recall} \quad (4)$$

TP (True Positive) represents the number of correctly classified anomalies, TN (True Negative) represents normal log entries and FP (False Positive) is the number of incorrect anomaly classification. FN (False Negative) is the number of incorrect classifications of log entries as normal while the label or ground truth states otherwise.

*C. Experimentation Preparation and Evaluation*

We populated the vector database using two approaches. The first approach was to populate the database with randomly selected samples from the log datasets that contain only normal log entries. The second approach was to populate the database with selected samples of the log database with normal log entries. For this selection, we first applied unsupervised k-means clustering to the dataset and populated the database from random sampling from the cluster classes. We used the elbow approach to select the number of cluster classes. For both approaches, to facilitate our evaluation, we kept the same number of records persisted in the vector databases.

While we applied the same evaluation techniques for both log datasets, we observed notable differences in the distribution of log patterns for both: namely, BGL has a wider distribution surface compared to Thunderbird. The following are the k-means clustering visualizations of both datasets.

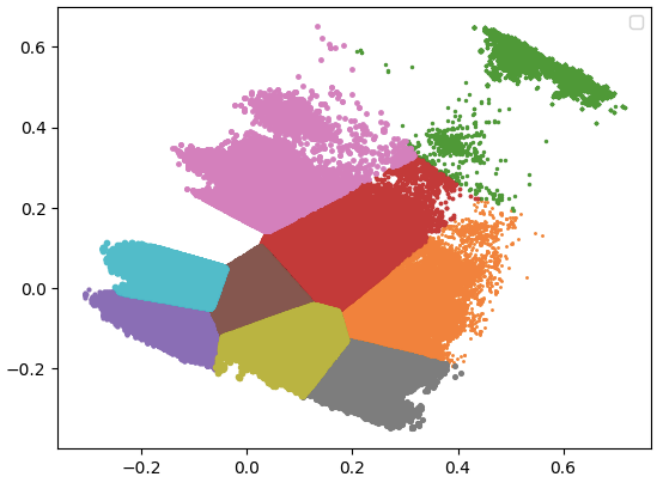

Chart 1. BGL Cluster Visualization Map

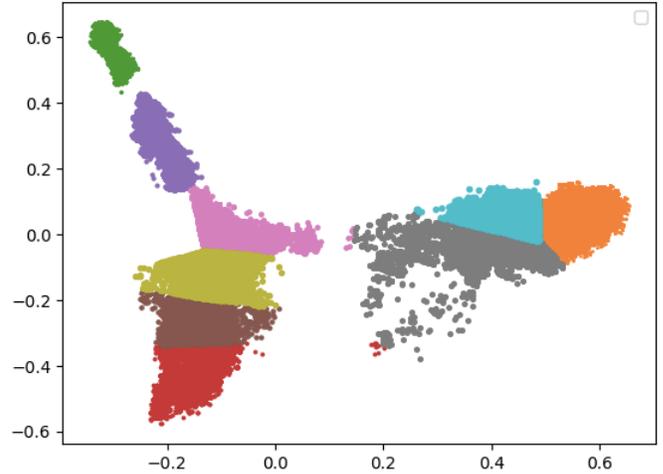

Chart 2. Thunderbird Cluster Visualization Map

For our experiment, we configured the sample size of 10,000 log entries for each cluster class and cluster classes of 5 for BGL dataset. For random sampling, that would be 50,000 log entries that are randomly selected. For Thunderbird, we used the same sample size of 10,000 from each of the 4 classes for the clustered approach and 40,000 for random selection.

After selecting log entries samples, they were then populated into the vector database. Using one predefined prompt template from a Question and Answer pipeline, we will assess the efficacy of our solution to detect log anomalies. In our prompt template, we explicitly directed GPT 3.5 (Davinci) with temperature of 0.1 to generate answers in the form of 'normal' or 'abnormal' to facilitate our evaluation. Due to cost constraints of using GPT 3.5, we randomly sampled from the 20% of both log datasets that has been set aside for testing.

*D. Results and Analysis*

From our experiment test results shown below in Chart 3, we observed that the clustering approach yielded better results for BGL log datasets as compared to random selection.

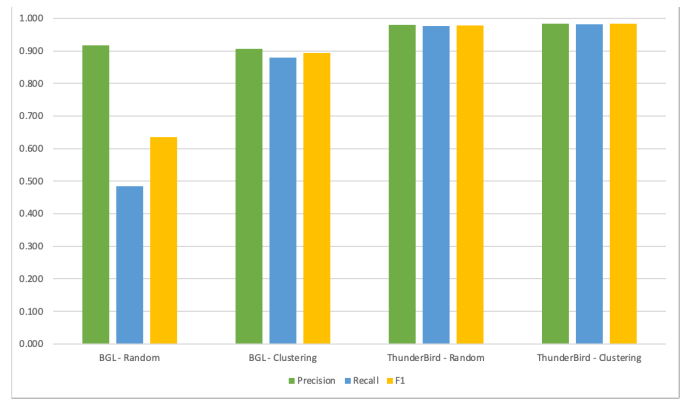

Chart 3. Evaluation performance comparison between clustered approach and random selection for both datasets.

The Thunderbird log dataset generally performed well with both randomized and clustering approach. This could be due to concentration of the log pattern distribution for this dataset.

We further evaluated our results with others who had applied zero shot classification using LLM [20]. Also the output from the LLM had either normal or abnormal returns with no other textual hallucination noted.

|  | Precision | Recall | F1 score |
|---|---|---|---|
| LogPrompt [20] | 0.25 | 0.83 | 0.38 |
| RAGLog (Ours) | 0.91 | 0.88 | 0.89 |

Table 1. Evaluation Comparison for Zero-Shot Classification

## VI. Conclusion and Future Directions

Our research work explored the use of Retrieval Augmented Generation model as log anomaly detector (RAGLog). The model's vector database only contained samples of normal log entries that were selected using unsupervised k-means clustering. It achieved good F1 scores when analyzing log entries using zero shot approach for anomalies, with the LLM being given only normal log entries for semantic analysis.

The constraints posed by this approach is the high resource consumption and execution latency for running the LLM and performing log analysis one log entry at a time. Hence, our next step will be to further optimize our RAG model approach to analyze logs faster with larger volumes.